\documentclass{aa}
\input psfig.sty
\newcommand{\nat}[2]{Nat #1, #2}
\newcommand{\apj}[2]{ApJ #1, #2}

\newcommand{\aj}[2]{AJ #1, #2}

\newcommand{\aeta}[2]{A\&A #1, #2}

\newcommand{\aetas}[2]{A\&AS #1, #2}

\newcommand{\nh}{N$_{\rm H}$}

\newcommand{\ergs}{erg s$^{-1}$}
\newcommand{\kms}{km s$^{-1}$}

\newcommand{\Msol}{M$_{\odot}$ }
\newcommand{\Teff}{$T_{\rm eff}$\,}
\newcommand{\Tbb}{$T_{\rm bb}$\,}

\newcommand{\degree}{\degr}

\newcommand{\rx}{\object{RX\,J0720.4-3125}}
\newcommand{\rxx}{\object{RX\,J1856.5-3754}}
\begin{document}

   \thesaurus{06         
              (13.25.3;  
               08.14.1;}  
   \title{Constraints on optical emission from the isolated neutron star candidate \rx
\thanks{Based on observations obtained at the European Southern Observatory, 
La Silla (Chile) with the NTT and ESO-Dutch telescopes}
    }

   \titlerunning{Constraints on optical emission from \rx }
 
   \author{C. Motch \inst{1}
       \and    F. Haberl   \inst{2}
          }

   \institute{
              Observatoire Astronomique, UA 1280 CNRS, 11 rue de l'Universit\'e, 
              F-67000 Strasbourg, France
              \and 
              Max-Planck-Institut f\"ur extraterrestrische Physik, D-85740,
              Garching bei M\"unchen, Germany 
              }
 
   \date{Accepted for publication in Astronomy \& Astrophysics, Letters}

   \offprints{C. Motch}

   \maketitle

   \begin{abstract}

Deep optical B band images of the ROSAT HRI error region of \rx \ reveal the presence
of two faint stellar-like objects with B = 26.1 $\pm$ 0.25 and B = 26.5 $\pm$ 0.30. 
Exposures obtained through U, V and I filters are not sensitive enough to detect the
two candidates and provide upper limits of U = 24.9, V = 23.2 and I = 21.9.  These new
observations virtually establish that \rx \ is a slowly rotating, probably completely
isolated neutron star. The absence of an optical counterpart brighter than B = 26.1
seems incompatible with a neutron star atmosphere having a chemical composition
dominated by Hydrogen or Helium.  UBI photometry of field stars shows astonishingly
little interstellar reddening in the direction of the X-ray source. Together with the
small column density detected by the ROSAT PSPC, this suggests a mean particle density
in the range of $n$ = 0.1 - 0.4 cm$^{-3}$. Such average densities would imply very low
velocities relative to interstellar medium ($v_{rel}$ $\leq$ 10\,\kms ) if the source
were powered by accretion.  These stringent constraints may be relaxed if the neutron
star is presently crossing a small size structure of higher density or if the effective
temperature of the heated atmosphere is overestimated by the blackbody approximation.
Alternatively, \rx \ could be a young and highly magnetized cooling neutron star.  


   \end{abstract}

%

\section{Introduction}

Since the pioneering work of Ostriker et al. (1970), several studies have demonstrated
that isolated old neutron stars (IONS) accreting matter from the interstellar medium
sh\-ou\-ld show up in great numbers at X-ray energies. As their expected effective
temperatures are in the range of 10 to 300 eV, with soft X-ray luminosities of up to
10$^{32}$\,\ergs \ this population should be outstandingly visible in ROSAT survey and
pointing data.  

In spite of the large number (10$^8$ to 10$^9$) of neutron stars created during the
past life of the Galaxy, their identification in ROSAT data has been so far elusive. 
The optical emission from such objects may only arise from the small surface heated by
accretion.  With most telescopes currently available, the faint counterparts (V $\geq$
25) cannot be recognized as such and accordingly, the density of IONS may only be
constrained from the number of unidentified ROSAT sources.  Optical campaigns carried
out in selected ROSAT fields at low and high galactic latitudes yielded an unexpectedly
small number of possible candidates (Motch et al.  1997, Zickgraf et al.  1997) which
constrained the IONS population space density to be more than a factor 10 lower than
predicted by current models (e.g.  Blaes \& Madau \cite{bm93}).

However, a couple of bright soft X-ray sources were successfully associated with
isolated neutron stars and in the X-ray brightest case (\rxx ) an HST observation
revealed the faint (V = 25.6) blue counterpart (Walter et al.  \cite{walter97}). The
second X-ray brightest candidate, \rx , exhibits the interesting feature to pulsate at
a period of 8.39s (Haberl et al. \cite{haberl97}). The short rotational period implies
a surface magnetic field of less than 10$^{10}$G if the source is powered by accretion.
This could provide the first good evidence for secular magnetic field decay in neutron
stars (Haberl et al. \cite{haberl97}, Wang \cite{wang97}) unless the decay results from
a past episode of intense accretion occurred in a yet destroyed binary system.  

\section{Optical Observations}

Optical observations were carried out using the ESO-NTT in service mode for the U and B
bands and the ESO-Dutch 0.9\,m telescope for the V and I bands. 

At the NTT we used SUSI equipped with the 1K$^{2}$ Tektronix TK 1024A CCD\#42. The
24$\mu$ pixel size corresponds to 0.13\,\arcsec \ on the sky. During the time interval
from 1997 February 7 till April 4 we accumulated 2.5\,h of observations in B and
3.5\,h in U. Raw images were corrected for bias using over-scan regions and
flat-fielded using sky images obtained at dawn. Individual 10\,min long exposures were
then stacked using a statistical method which rejected cosmic ray impacts. Finally, as
the average FWHM seeing was about 0.9\,\arcsec \ merged images were rebinned to a
0.39\,\arcsec \ pixel size in order to gain sensitivity.  Absolute photometric
calibration and colour transformation was derived using standard fields from Landolt
(\cite{landolt}) observed during four consecutive nights in February 1997.  

At the ESO-Dutch 0.9\,m telescope, we used the standard CCD adaptor equipped with
Tektronics TK 512 CCD \#33. This chip has a pixel size of 27$\mu$, corresponding to
0.44\,\arcsec \ on the sky. A total of 17\,min of observing time was accumulated in the
V band on 1997 February 9 and 25\,min in I band on February 11. CCD frames were
corrected applying the same methods as for NTT data. The \object{PG1323} field (Landolt
\cite{landolt}) was used for photometric calibration.

\section{Optical content of the ROSAT HRI error circle}

Following Haberl et al. (\cite{haberl97}) we consider two possible boresight corrected
HRI positions. Both positions are consistent one with each other and have each a 90\%
confidence radius of 3\,\arcsec.
 
In the U, V and I band images, we fail to detect any object in or near the ROSAT error
circles. We estimated the upper limit on the brightness of any source by generating at
the ROSAT position artificial stellar images with a point spread function derived
from a bright non-saturated stellar profile. Any star brighter than U = 24.9, V = 23.2
and I = 21.9 would have been detected.

\begin{figure}
\psfig{figure=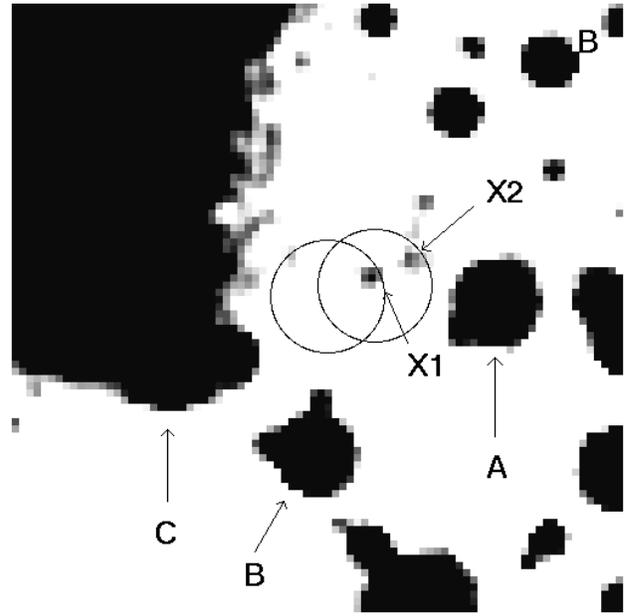,width=8.8cm,bbllx=2.0cm,bburx=19cm,bblly=10.0cm,bbury=27cm,clip=true}
\caption[]{ROSAT HRI error circles overlayed on the B image smoothed with a Gaussian
filter ($\sigma$ = 0.35\arcsec). The size of the image is about 33\arcsec. 
North is to the top and east to the left}
\label{bfc}
\end{figure}

The B band image reveals the presence of two objects, X1 and X2 in the merged error
circles (see Fig. \ref{bfc}). Object X1 has B = 26.1$\pm$0.25 and X2, B =
26.5$\pm$0.30.  The upper limit on the brightness of any other star is B = 26.5.  

\section{ISM properties towards \rx}

\begin{figure}
\psfig{figure=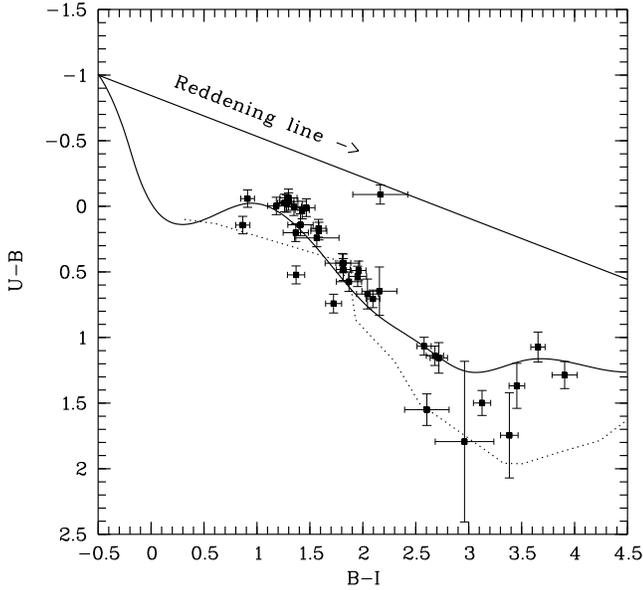,width=8.8cm,bbllx=1.0cm,bburx=20cm,bblly=10.0cm,bbury=28cm,clip=true}
\caption{U-B/B-I colour diagram for 37 stars located within 1\arcmin \ from the HRI position of
\rx. The relations for main sequence stars (solid line) and giant (dotted line) 
are plotted for comparison} 
\label{phot} 
\end{figure}

We show on Fig.  \ref{phot} the U-B / B-I diagram for 37 faint objects located within
1\arcmin \ from the ROSAT HRI position together with mean relations for main sequence
and evolved stars. Errors on individual colour indices take into account a 0.05 mag
systematic uncertainty on the colour transformation. Because of CCD saturation we
could not measure any of the bright stars located NE of the ROSAT position. B
magnitudes are in the range of 17.8 - 23.6 with a peak between 20 and 23.  Most
objects have colour indices and magnitudes compatible with main sequence stars
undergoing very little interstellar absorption. At the red end, few stars have colours
suggesting some evolution. The couple of objects significantly departing from the main
sequence relation may be extragalactic objects. UBV photometry of objects A, B and C
(see Table \ref{abc}) confirms the classification proposed by Haberl et al.
(\cite{haberl97}) on the basis of optical spectroscopy. A and B are G0 and G5 main
sequence type stars whereas the colours of C suggest a K3 giant rather than a M dwarf
type star.  

\begin{table}
\caption{UBV photometry of objects A,B and C}
\label{abc}
\begin{tabular}{cccc}
Star        & V   & B-V  &  U-B   \\ \hline
A           & 20.00$\pm$0.05 & 0.57$\pm$0.07 & 0.00$\pm$0.07 \\
B           & 20.13$\pm$0.05 & 0.70$\pm$0.07 & 0.17$\pm$0.07 \\
C           & 19.91$\pm$0.05 & 1.32$\pm$0.07 & 1.50$\pm$0.10 \\            
\hline
\end{tabular}
\end{table}

The group of 15 stars with U-B $\leq$ 0.3 and B-I $\leq$ 2.0 are the most luminous
unevolved objects in the field and probe absorption up to a mean distance of 14\,kpc,
i.e, well outside the galactic disc.  By fitting a reddened population I main sequence
relation to this group of stars we derive a most probable E(B-V) excess of 0.08 and
put a 95\% confidence upper limit of E(B-V) = 0.12 or \nh = 7.0 10$^{20}$ H atom
cm$^{-2}$.  As a large fraction of these high $z$ stars may be low metallicity
population II subdwarfs, we also fitted a U-B / B-I relation assuming $[$ Fe / H $]$ =
$-$1.0 and using calibrations provided by Cameron (\cite{cam85}). The population II
relation gives a better fit to the photometry with a best E(B-V) excess of 0.04 and a
95\% confidence upper limit of E(B-V) = 0.14.  The group of 3 main sequence stars with
B-I comprised between 2.5 and 2.8 and located at $\sim$ 2.2\,kpc does not provide
useful additional constraint at smaller distances.

In this direction the total galactic HI column density averaged over 0.5\degree \ is
1.89 10$^{21}$ H atom cm$^{-2}$. However, the IRAS 10$\mu$ map shows a patchy
structure with a hole only 4\arcmin \ SW to \rx \ suggesting that the actual \nh \
towards \rx \ may well be as low as indicated by our field star photometry. 
Therefore, both the column density to the source, \nh \ = 1.3$\pm$0.3 10$^{20}$ H atom
cm$^{-2}$ derived from X-ray spectral fitting and that from the source to the edge of
the Galaxy are likely to be low.  In particular, there is no evidence for any dense
cloud behind the X-ray source as in the case of RX\,J1856.5-3754 (Walter et al.
\cite{walter96}).  

If the $z$ distribution of HI along the direction of the X-ray source has the general
shape of that given in Dickey \& Lockman (\cite{dickey}) then a total \nh \ of 7.0
10$^{20}$ H atom cm$^{-2}$ would imply an average particle density of $\sim$ 0.2
cm$^{-3}$ at distances of a few hundred pc.  A tenuous interstellar medium of
comparable low mean density ($n$ $\sim$ 0.2 cm$^{-3}$) is also found towards the
nearby (440\,pc) open cluster \object{Collinder 140}, located only 1\degree \ away
(Clari\`a \& Rosenzweig \cite{claria}). Finally, \rx \ is less than 5\degree \ away
from the interstellar tunnel of neutral free gas discovered by Welsh (\cite{welsh}).
This rarefied region which has a diameter of 50\,pc and a length of 300\,pc is thought
to be the farthest extension of the Local Bubble of hot gas.  

PSPC data also imply $n$ = 0.4 ($d$/100pc)$^{-1}$ cm$^{-3}$ with $d$
approximately in the range of 100 to 440\,pc (Haberl et al.  \cite{haberl97}). 
Therefore, averaged over a large scale length, the ambient interstellar medium density
is $n$ = 0.1 - 0.4 cm$^{-3}$, significantly smaller than the mean galactic plane value
($n$ = 0.57 cm$^{-3}$, Dickey \& Lockman \cite{dickey}).  
 
\section{Discussion}

\subsection{Is \rx \ a lonely neutron star ?}

The first important piece of information provided by the NTT observations is the
absence of an optical counterpart brighter than B = 26.1 in the HRI error circles. This
low optical flux definitely rules out any other kind of identification than with a
neutron star.  At a maximum distance of 440\,pc, constrained by the blackbody fit to the
PSPC energy distribution, an hypothetical companion to the neutron star would have an
absolute I magnitude fainter than 13.7. Only stars later than M7V and less massive than
0.09 \Msol may be faint enough to remain undetected. At 100\,pc the hypothetical
companion can only be a brown dwarf with M $\sim$ 0.075 \Msol (Baraffe \& Chabrier
\cite{ba}). The low mass companion star would not have a wind intense enough to fuel
the neutron star and the only mechanism which would allow the required mass transfer
rate ($\dot{\mathrm{M}}$ $\sim$ 1.2 10$^{11}$ g\,s$^{-1}$, $d$ = 100\,pc) is Roche lobe
overflow. In such a case, we would expect substantial X-ray heating of the brown dwarf
atmosphere and again an absolute I magnitude brighter than our limit.  Therefore, we
can probably conclude that the compact object is either completely isolated or at most
accompanied by a very late M star or brown dwarf companion and that in all cases the
neutron star does not accrete from the companion star.

\subsection{Did we detect the optical counterpart of \rx ?} 

Extrapolating the blackbody detected at soft X-ray energies into the optical regime
indicates that the neutron star should appear as a very faint but extremely blue
object (B = 28.2, U-B = $-$ 1.3), 1.7 magnitude fainter than our B upper limit. 
However, there are several observational and theoretical arguments in favour of a
brighter optical flux from \rx .  First, for a given soft X-ray flux, the optical
magnitude depends very sensitively upon the chemical composition of the neutron star
atmosphere.  Rajagopal \& Romani (\cite{raro}) and Zavlin et al. (\cite{zavlin}) have
computed model atmospheres of low magnetic field neutron stars for various chemical
compositions from pure Fe to pure H.  These authors show that blackbody spectral fits
to ROSAT PSPC data may yield temperatures and bolometric luminosities very different
from those of model atmospheres. Atmospheres dominated by Hydrogen or Helium exhibit
rather hard spectra in the 0.1 - 1.0 keV range which folded through the ROSAT PSPC and
fitted by blackbody models give temperatures up to 3 times larger than the actual
effective temperature of the neutron star. Similar effects albeit of lower amplitudes
apply to a pure iron atmosphere. For instance, \Tbb = 79 eV (Haberl et al. 
\cite{haberl97}) would correspond to \Teff = 63 eV for pure iron atmosphere and only
\Teff = 27 eV for Hydrogen dominated atmospheres (Rajagopal \& Romani \cite{raro}). 
Furthermore, optical flux may be 1.5 to 5 magnitude above blackbody level, strongly
depending on chemical composition (Pavlov et al. \cite{pavlov97}, Walter et al.,
\cite{walter97}). Our limit of B = 26.1 apparently rules out H or He dominated
atmospheres.

On the observational side, Walter et al. (\cite{walter97}) report the HST discovery of
the optical counterpart of the related source \rxx \ with a V = 25.6 magnitude blue
excess object (U-V = $-$ 1.2). The optical continuum is a factor 3.7 (1.4 mag) above
the extrapolation of the ROSAT blackbody at 606 nm. Although the young pulsar physical
conditions of \object{Geminga} may not allow direct comparison with \rx , significant
discrepancy between optical flux level and expected Rayleigh-Jeans continuum has been
observed by Bignami et al. (\cite{big}). Therefore, it may well be that we already have
enough sensitivity in B to detect the optical emission from the surface of the neutron
star and that object X1 (B = 26.1) or X2 (B=26.5) is the counterpart of \rx. 
Unfortunately, the V and U magnitude limits are not deep enough to yield useful
constraints on the colour indices of these candidates.

\subsection{Is \rx \ powered by accretion from ISM or by neutron star cooling ?}

One intriguing feature of \rx \ is the low mean particle density of the medium in which
the neutron star travels. There is some kind of contradiction in finding the second
X-ray brightest isolated neutron star, believed to be powered by accretion from
interstellar medium, in one of the lowest density regions close to the Sun. Assuming
blackbody emission and Bondi-Hoyle accretion, $v_{rel}$, the relative velocity of \rx \
with respect to interstellar medium is given by:

\begin{displaymath}
(v_{rel}^{2} + c_{s}^{2})^{\frac{1}{2}}\ = 
\ 16.7\, n^{\frac{1}{3}}\, d_{100}^{-\frac{2}{3}} \ {\mathrm km\,s^{-1}}
\end{displaymath}

with $ c_{s}$ the sound velocity, $n$ the hydrogen number density of the interstellar
medium and $d_{100}$ the distance to the source in units of 100\,pc. If the sound
velocity is of the order of 10 \kms \ (e.g. Blaes \& Madau \cite{bm93}), and $n$ in the
range of 0.1 - 0.4, then $v_{rel}$ must be amazingly low to account for the observed
blackbody temperature derived from fits to PSPC data. In fact, the relative velocity
reaches null values already at $d$ = 140\,pc ($n$ = 0.4 cm$^{-3}$) and as close as $d$
= 70\,pc ($n$ = 0.1 cm$^{-3}$).  

HI structures are known to exist on scales ranging from 1\,kpc to less than 100 AU
(Dickey \& Lockman \cite{dickey}) and the particle density may be locally much larger
or smaller than on average. Hot bubbles such as our Local Bubble can generate high
density shells. The contributions of these dense regions to the total \nh \ may be too
small to be detected by optical reddening measurements while offering particle
densities in excess of 1 cm$^{-3}$ on distances of 10\,pc or more. A similar
consideration may be applied to even smaller structures.  Cloudlets with size of 1000
UA and densities of 10$^{3}$ cm$^{-3}$ seem ubiquitous in the Galaxy (Watson \& Meyer
\cite{wm96}). One could thus solve the density problem by supposing that \rx \ is now
crossing the Local Bubble boundary or is immersed in one of these high density
cloudlets. Long term variations of the X-ray luminosity could then soon give
information on the size of the interstellar cloud surrounding the X-ray source.  If the
accretion luminosity is lower than estimated from the blackbody fit, the stringent
constraints on relative velocity may be relaxed even in the case of a low density.

Alternatively, \rx \ could be a young cooling neutron star as proposed by Heyl \&
Hernquist (\cite{hh}), a model which would not require any accretion at all. In this
case, a very high polar field of $\sim$ 10$^{14}$G must be present in order to slow
down the neutron star by magnetic dipole radiation in only $\sim$ 3 10$^{5}$ yr.

\begin{acknowledgements}
We thank Albert Zijlstra and the ESO staff for preparing and conducting the NTT
observations. We are grateful to the referee, F. Walter, and to J. Tr\"umper for
helpful comments and discussions. The ROSAT project is supported by the German Bundesministerium
f\"ur Bildung, Wissenschaft, For\-schung und Technologie (BMBF/DLR) and the
Max-Planck-Gesellschaft. 
\end{acknowledgements}


\begin{thebibliography}{}

   \bibitem[1996]{ba}
    Baraffe, I., Chabrier, G., 1996, \apj{461}{L51}

   \bibitem[1996]{big}
    Bignami, G.F., Caraveo, P.A., Mignami, R., Edelstein, J., Bowyer, S., 1996,
    \apj{456}{L111}

   \bibitem[1993]{bm93} 
    Blaes, O., Madau, P., 1993, \apj{403}{690}

   \bibitem[1985]{cam85}
    Cameron, L. M., 1985, \aeta{146}{59}
 
   \bibitem[1978]{claria}
    Clari\`a, J.J., Rosenzweig, P., 1978, \aj{83}{278}

   \bibitem[1990]{dickey}
    Dickey, J., Lockman, F., 1990, ARA\&A, 28, 215

   \bibitem[1997]{haberl97} 
    Haberl, F., Motch, C., Buckley, D.A.H., Zickgraf, F.-J., Pietsch, W., 1997, 
    \aeta{326}{662} 
    
   \bibitem[1998]{hh}
    Heyl, J.S., Hernquist, L., 1998, preprint

   \bibitem[1992]{landolt} 
    Landolt, A.U., 1992, \aj{104}{340}

   \bibitem[1997]{motch97} Motch, C., Guillout, P., Haberl, et al., 1997,
   \aeta{318}{111}

   \bibitem[1997]{pavlov97}
   Pavlov, G.G., Zavlin, V.E., Tr\"umper, J., Neuh\"auser, 1997, \apj{472}{L53}

   \bibitem[1996]{raro} Rajagopal, M., Romani, R., 1996, \apj{461}{327}

   \bibitem[1997]{wang97} Wang, J.C.L., 1997, \apj{486}{L119} 

   \bibitem[1996]{walter96}
   Walter, F.M., Wolk, S.J., Neuh\"auser, R., 1996, \nat{379}{233}

   \bibitem[1997]{walter97} 
    Walter, F. M., Matthews, L.D., Lattimer, P. An. J., Prakash, M., 
Neuh\"auser, R., 1997, Proceedings of the 1997 HEAD meeting, Nov 4-7 1997, in press

   \bibitem[1996]{wm96}
    Watson, J. K., Meyer, D. M., 1996, \apj{473}{L127}  

   \bibitem[1991]{welsh}
    Welsh, B.Y., 1991, \apj{373}{556} 

   \bibitem[1996]{zavlin}
    Zavlin, V.E., Pavlov, G.G., Shibanov, Yu.A., 1996, \aeta{315}{141}

   \bibitem[1997]{z97} Zickgraf, F.-J., Thiering, I., Krautter, J., et al.,  
   1997, \aetas{123}{103} 

\end{thebibliography}
\end{document}